\documentclass[sigconf]{acmart}
\usepackage{booktabs}
\AtBeginDocument{%
  }

\copyrightyear{2025}
\acmYear{2025}
\setcopyright{rightsretained}
\acmConference[AHs '25]{The Augmented Humans International Conference 2025}{March 16--20, 2025}{Masdar City, Abu Dhabi, United Arab Emirates}
\acmBooktitle{The Augmented Humans International Conference 2025 (AHs '25), March 16--20, 2025, Masdar City, Abu Dhabi, United Arab Emirates}
\acmPrice{}
\acmDOI{10.1145/3745900.3746107}
\acmISBN{979-8-4007-1566-2/25/03}
\begin{document}

\title{Look and Talk: Seamless AI Assistant Interaction with Gaze-Triggered Activation}


\author{Qing Zhang}
\orcid{0000-0002-0622-932X}
\affiliation{%
  \institution{The University of Tokyo}
  \city{Tokyo}
  \country{Japan}
}
\email{qzkiyoshi@gmail.com}

\author{Jun Rekimoto}
\orcid{0000-0002-3629-2514}
\affiliation{%
  \institution{The University of Tokyo}
    \institution{Sony CSL - Kyoto}
  \city{Tokyo, Kyoto}
  \country{Japan}}
\email{rekimoto@acm.org}

\renewcommand{\shortauthors}{Zhang et al.}

\begin{abstract}
  Engaging with AI assistants to gather essential information in a timely manner is becoming increasingly common. Traditional activation methods, like wake words such as Hey Siri, Ok Google, and Hey Alexa, are constrained by technical challenges such as false activations, recognition errors, and discomfort in public settings. Similarly, activating AI systems via physical buttons imposes strict interaction limitations as it demands particular physical actions, which hinders fluid and spontaneous communication with AI. Our approach employs eye-tracking technology within AR glasses to discern a user's intention to engage with the AI assistant. By sustaining eye contact on a virtual AI avatar for a specific time, users can initiate an interaction silently and without using their hands. Preliminary user feedback suggests that this technique is relatively intuitive, natural, and less obtrusive, highlighting its potential for integrating AI assistants fluidly into everyday interactions.
\end{abstract}


\begin{teaserfigure}
  \includegraphics[width=\textwidth]{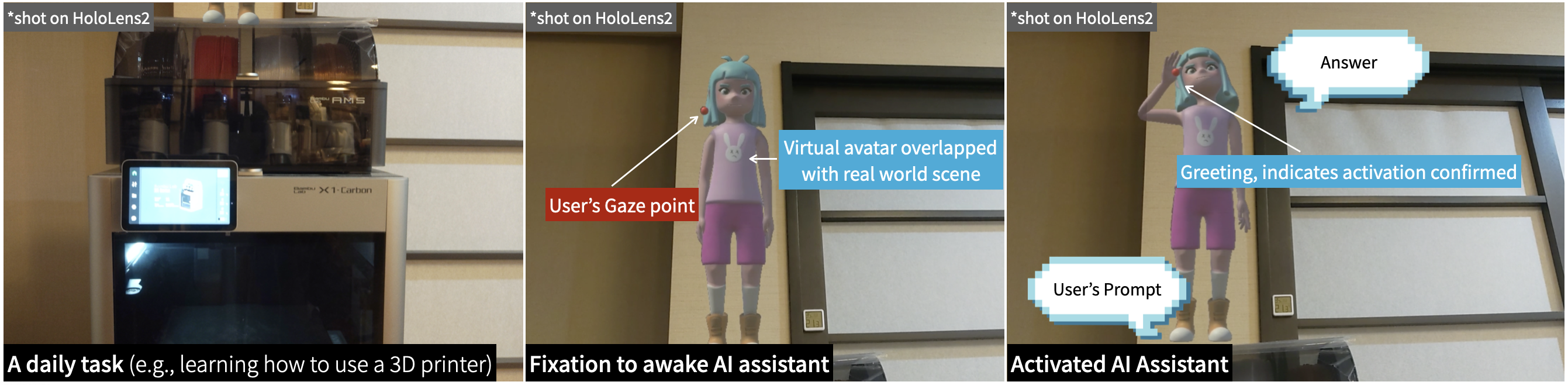}
  \caption{A gaze-triggered AI assistant activation system using optical see-through AR headset. Left: User engaged in a daily task with the virtual avatar positioned peripherally. Center: When the user directs their gaze at the virtual avatar (fixation point), the system detects potential interaction intent. Right: After maintaining gaze fixation for two seconds, the AI assistant activates, confirming with a greeting animation and awaiting the user's voice input without requiring wake words or physical controls.}
  \Description{Enjoying the baseball game from the third-base
  seats. Ichiro Suzuki preparing to bat.}
  \label{fig:teaser}
\end{teaserfigure}

\maketitle

\section{Introduction}
    The increasing prevalence of AI agents in daily life, from virtual assistants on smartphones to smart eyewear (such as Google Glass and Ray-Ban Meta), has made interacting with these agents a common experience. Especially, experimental prototypes such as Meta Aria and Orion forecast that AI-powered everyday eyewear could equip eye tracking and optical see-through displays without being bulky at all. However, current methods for initiating interaction with AI assistants often fall short of seamless and natural communication. Voice-activated systems, triggered by ``\textit{wake words}'' like ``Hey Siri'' or ``Ok Google,'' are prone to issues like accidental activation (false positives) \cite{ahmed2022towards}, failure to recognize the wake word (false negatives), and social awkwardness in public settings. Physical button activation, while reliable, introduces a physical constraint and interrupts the flow of interaction, requiring the user to explicitly break their focus from their ongoing work to engage with the assistant. These limitations hinder the seamless interaction of the AI assistant in daily usage \cite{lemay2021virtual}. 
    
    The most recent research has explored various approaches to address these interaction challenges. Voice interaction remains the dominant paradigm, with ongoing efforts to shorten the wake words and improve robustness and accuracy \cite{9to5macSiriLonger}. In terms of research attempts, Mayer et al. \cite{mayer2020enhancing} explored a worldgaze-based approach using a smartphone's front and a rear camera to equip the voice assistant with context awareness. Similarly, lee et al. \cite{lee2024gazepointar} utilized Hololens2 for a spatially aware system in human voice assistant communication. Despite advances in previous work, there is still a gap in initiating AI assistant interaction within the daily human-AI interaction. Current approaches either require explicit voice commands that can be socially disruptive or demand physical actions that interrupt ongoing tasks. What is missing is an interaction method that feels natural and unobtrusive while maintaining user control over activation.

    This research proposes a novel approach to AI assistant interaction in an everyday scenario that leverages user eye fixation as a signal of intent to initiate the AI assistant. Our system, built using a Microsoft HoloLens 2, renders a virtual AI agent avatar above the user's line of sight, which is just outside the visual field. The HoloLens 2's built-in eye-tracking capabilities continuously monitor the user's gaze direction. A dwell-time threshold (two seconds) is used to distinguish between intentional gaze and casual glances; when the user's gaze remains focused on the AI agent's avatar for a predetermined duration, the system interprets this as an intention to interact. Upon detecting intentional gaze, the agent greets the user by waving hands, confirming recognition, and entering an ``interaction mode,'' where the user can naturally talk to the assistant.

    The preliminary user test has yielded encouraging results. The participants reported that the fixation-based initiation of the assistant was intuitive and natural, appreciating the silent and hands-free nature of the activation. Feedback highlighted the importance of carefully tuning the dwell time threshold and the acknowledgment feedback design. These initial findings suggest that this approach holds potential for creating more seamless and less intrusive interactions with AI assistant in smart AR eyewear that is equipped every day.

\section{Related Works}

Research in human-computer interaction has explored various modalities for initiating and interacting with AI systems, which can be categorized into several key areas.

\textbf{Voice-Based Activation: }Traditional voice assistants rely heavily on wake-up words such as ``Hey Siri'' or ``Ok Google'' for activation. Although these systems offer hands-free operation, they face significant challenges, including false triggers, recognition failures, and social awkwardness in public settings \cite{chen2021fakewake}. Recent developments have focused on shortening wake words and improving robustness \cite{9to5macSiriLonger, ahmed2022towards}, but fundamental limitations in voice-based activation persist. These include privacy concerns in public settings, high false activation rates in noisy environments, and the social discomfort of verbally addressing technology. Such limitations highlight the need for more natural, unobtrusive activation methods that seamlessly integrate into users' daily activities, especially for wearable AR devices.

\textbf{Gaze-Based Interaction: }Eye tracking has emerged as a promising modality for human-computer interaction, particularly in XR (mixed reality) environments. Recent work has demonstrated that gaze data can be effectively used to infer user intentions and cognitive states \cite{yuan2021invoking}. However, most existing research focuses on using gaze for direct selection or control rather than as an activation mechanism for AI agents.

\textbf{Gesture Recognition: }Some systems utilize hand gestures or body movements for device control, particularly in XR environments \cite{oudah2020hand}. Although gesture-based interfaces can be intuitive, they often require explicit and sometimes exaggerated movements, which can be physically demanding and socially conspicuous.

\textbf{Context-Aware Systems: }Recent research has explored the combination of multiple interaction modalities to create more context-aware systems. Mayer et al. developed a worldgaze approach using smartphone cameras to provide context awareness to voice assistants \cite{mayer2020enhancing}. Similarly, Lee et al. utilized HoloLens 2 for spatial awareness in human-AI communication \cite{lee2024gazepointar}. These systems demonstrate the potential of multimodal interaction, but still rely primarily on voice commands for activation.

\textbf{AI Agent Interaction in XR: }The integration of AI agents in XR environments has gained increasing attention. Recent work has explored various approaches to make such interactions more natural and intuitive. For instance, some researchers have investigated the use of explainable AI interfaces to improve user understanding and interaction quality in XR settings \cite{yu2024explainable}. However, the specific challenge of naturally initiating interaction with AI agents in XR environments remains relatively unexplored.

Our work builds on previous research while addressing its limitations. Using eye fixation as the primary trigger for interaction with AI assistants, we propose a more natural and less intrusive paradigm for everyday use. This approach is distinct because it focuses on silent, hands-free initiation, mirroring natural human communication patterns. Just as humans establish eye contact to signal readiness for conversation, our system uses sustained gaze as an intuitive indicator of a user's intention to engage with the AI assistant. 

\begin{figure*}
    \centering
    \includegraphics[width=\linewidth]{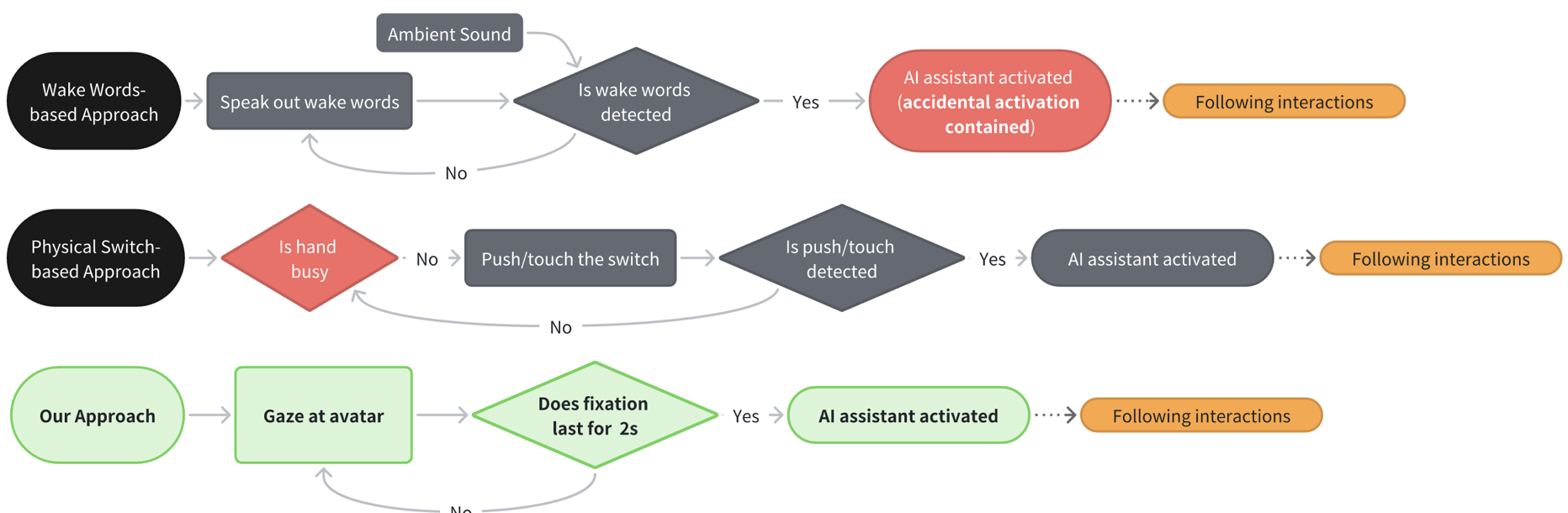}
    \caption{Comparison of AI assistant activation methods. Top: Wake word-based approach requires voice commands that may be affected by ambient noise and can lead to accidental activations. Middle: Physical switch-based approach requires deliberate manual interaction, which is problematic when hands are busy. Bottom: Our proposed gaze-based approach enables hands-free, silent activation through sustained eye fixation (2s threshold), providing a more intuitive and less intrusive interaction method while minimizing accidental triggers.}
    \label{fig:comparison}
\end{figure*}

As in the comparison among existing methods that are illustrated in Figure \ref{fig:comparison}, wake-word-based approaches rely on accurate speech recognition, which can be prone to errors, while physical switch activation requires deliberate action from the user, interrupting their workflow. In contrast, our gaze-based approach offers a more seamless and intuitive interaction utilizing eye fixation as the trigger.

\section{Our Approach: Eye Fixation As The Trigger of Virtual AI Assistant}
    Our methodology utilizes eye tracking to ascertain the user's gaze orientation and detects the interaction between where the user intends to direct their attention and a virtual avatar situated within a mixed reality environment. This approach is motivated by the typical initiation process in natural human communication, where an individual shifts their gaze from their current task to directly look at another person, facilitating conversation initiation and reducing misunderstandings. The virtual avatar is placed above the user's line of sight, nearly out of the visual field, with a small segment still visible to signal the position of the AI assistant (see Figure \ref{fig:teaser}). To activate the AI assistant, embodied by the virtual avatar, the user needs to fixate on the avatar for approximately two seconds. This duration was established through an informal test. This requirement effectively minimizes accidental activations due to brief glances. Upon activation, the user can verbally interact with the avatar without the necessity of using wake words or a physical switching mechanism.

\section{Initial Exploration}

    To evaluate our approach for preliminary insights, we created a prototype system using the Microsoft HoloLens 2, an optical see-through head-mounted display (HMD) equipped with integrated eye-tracking features. A virtual AI agent avatar was generated using Mixamo\footnote{\url{https://www.mixamo.com/}} and visualized in the HoloLens 2 environment. This study involved two participants—self-identified as one male and one female, aged 34 and 33 years, respectively. They were tasked with comparing interaction with a large language model-powered AI assistant using either our method or a traditional wake word-based approach. Considering delays in hardware and software as well as focusing on ideal scenario exploration, a \textit{wizard of oz} technique was utilized whereby an experimenter manually activated the AI assistant in response to the virtual avatar's activation, enabling participant-AI interaction. Approach order was counterbalanced, and each condition covered two to five basic tasks, such as requesting weather updates, current news, or addressing spontaneous queries.

\section{Preliminary User Feedback and Insights}

    Observations and informal interviews of the initial exploration revealed the following: (1) Naturalness: The participants generally found the fixation-based interaction to be intuitive and natural. Several commented that "I like the part of initiation with no need for wake words. Despite the advancement of AI assistants, each time when I orally wake up the AI assistant, I feel awkward no matter if there is another person present." (2) Quiet Activation and Hands-Free: Participants valued the eye-tracking activated AI assistant, especially within potentially loud or public settings. The ability to operate it without using hands was also regarded as a favorable feature. (3) Adjustment of dwelling time: Fine-tuning the dwell time threshold emerged as essential. Both participants (n=2) indicated that the duration of the activation was slightly excessive. This highlights the need for threshold optimization; a threshold too short could lead to inadvertent activation of the assistant, whereas an excessive duration may result in slow interaction. (4) Feedback Mechanism: One participant recommended implementing more prominent ``activation confirmed'' signals. This underscores the importance of noticeable feedback to boost user confidence in their commands. Incorporating a suggestion from a participant, who noted, ``Plus, wouldn't it be cooler if I could silently communicate with AI (assistant)?'', the integration of silent speech capabilities \cite{rekimoto2021derma, igarashi2022silent} could enhance the efficiency of Human-AI-Assistant interactions.
    
    This initial feedback might suggest that fixation-based initiation could be a less intrusive way to engage with AI assistants, particularly when AI-powered AR glasses become commonplace. Future research will aim to conduct in-depth user studies, explore diverse interaction modalities, and incorporate context awareness. We believe that this study has the potential to enhance human-AI interaction frameworks within the rapidly developing field of augmented reality, addressing the limitations of conventional voice and button interfaces by providing a quiet, hands-free, and more natural interaction method.

    

\begin{acks}
This work was supported by JST Moonshot R\&D Grant JPMJMS2012, and JPNP23025 commissioned by the New Energy and Industrial Technology Development Organization (NEDO).
\end{acks}

\bibliographystyle{ACM-Reference-Format}
\bibliography{sample-base}


\begin{thebibliography}{11}


\ifx \showCODEN    \undefined \def \showCODEN     #1{\unskip}     \fi
\ifx \showISBNx    \undefined \def \showISBNx     #1{\unskip}     \fi
\ifx \showISBNxiii \undefined \def \showISBNxiii  #1{\unskip}     \fi
\ifx \showISSN     \undefined \def \showISSN      #1{\unskip}     \fi
\ifx \showLCCN     \undefined \def \showLCCN      #1{\unskip}     \fi
\ifx \shownote     \undefined \def \shownote      #1{#1}          \fi
\ifx \showarticletitle \undefined \def \showarticletitle #1{#1}   \fi
\ifx \showURL      \undefined \def \showURL       {\relax}        \fi
\providecommand\bibfield[2]{#2}
\providecommand\bibinfo[2]{#2}
\providecommand\natexlab[1]{#1}
\providecommand\showeprint[2][]{arXiv:#2}

\bibitem[Ahmed et~al\mbox{.}(2022)]%
        {ahmed2022towards}
\bibfield{author}{\bibinfo{person}{Shimaa Ahmed}, \bibinfo{person}{Ilia Shumailov}, \bibinfo{person}{Nicolas Papernot}, {and} \bibinfo{person}{Kassem Fawaz}.} \bibinfo{year}{2022}\natexlab{}.
\newblock \showarticletitle{Towards more robust keyword spotting for voice assistants}. In \bibinfo{booktitle}{\emph{31st USENIX Security Symposium (USENIX Security 22)}}. \bibinfo{pages}{2655--2672}.
\newblock


\bibitem[Chen et~al\mbox{.}(2021)]%
        {chen2021fakewake}
\bibfield{author}{\bibinfo{person}{Yanjiao Chen}, \bibinfo{person}{Yijie Bai}, \bibinfo{person}{Richard Mitev}, \bibinfo{person}{Kaibo Wang}, \bibinfo{person}{Ahmad-Reza Sadeghi}, {and} \bibinfo{person}{Wenyuan Xu}.} \bibinfo{year}{2021}\natexlab{}.
\newblock \showarticletitle{Fakewake: Understanding and mitigating fake wake-up words of voice assistants}. In \bibinfo{booktitle}{\emph{Proceedings of the 2021 ACM SIGSAC Conference on Computer and Communications Security}}. \bibinfo{pages}{1861--1883}.
\newblock


\bibitem[Espósito(2023)]%
        {9to5macSiriLonger}
\bibfield{author}{\bibinfo{person}{Filipe Espósito}.} \bibinfo{year}{2023}\natexlab{}.
\newblock \bibinfo{title}{{S}iri no longer requires '{H}ey' command to activate --- 9to5mac.com}.
\newblock \bibinfo{howpublished}{\url{https://9to5mac.com/2023/06/05/siri-no-longer-requires-hey-command/}}.
\newblock
\newblock
\shownote{[Accessed 09-02-2025]}.


\bibitem[Igarashi et~al\mbox{.}(2022)]%
        {igarashi2022silent}
\bibfield{author}{\bibinfo{person}{Yuya Igarashi}, \bibinfo{person}{Kyosuke Futami}, {and} \bibinfo{person}{Kazuya Murao}.} \bibinfo{year}{2022}\natexlab{}.
\newblock \showarticletitle{Silent Speech Eyewear Interface: Silent Speech Recognition Method Using Eyewear with Infrared Distance Sensors}. In \bibinfo{booktitle}{\emph{Proceedings of the 2022 ACM International Symposium on Wearable Computers}}. \bibinfo{pages}{33--38}.
\newblock


\bibitem[Lee et~al\mbox{.}(2024)]%
        {lee2024gazepointar}
\bibfield{author}{\bibinfo{person}{Jaewook Lee}, \bibinfo{person}{Jun Wang}, \bibinfo{person}{Elizabeth Brown}, \bibinfo{person}{Liam Chu}, \bibinfo{person}{Sebastian~S Rodriguez}, {and} \bibinfo{person}{Jon~E Froehlich}.} \bibinfo{year}{2024}\natexlab{}.
\newblock \showarticletitle{GazePointAR: A Context-Aware Multimodal Voice Assistant for Pronoun Disambiguation in Wearable Augmented Reality}. In \bibinfo{booktitle}{\emph{Proceedings of the CHI Conference on Human Factors in Computing Systems}}. \bibinfo{pages}{1--20}.
\newblock


\bibitem[Lemay et~al\mbox{.}(2021)]%
        {lemay2021virtual}
\bibfield{author}{\bibinfo{person}{Stephen~O Lemay}, \bibinfo{person}{Brandon~J Newendorp}, {and} \bibinfo{person}{Jonathan~R Dascola}.} \bibinfo{year}{2021}\natexlab{}.
\newblock \bibinfo{title}{Virtual assistant activation}.
\newblock
\newblock
\shownote{US Patent 11,087,759}.


\bibitem[Mayer et~al\mbox{.}(2020)]%
        {mayer2020enhancing}
\bibfield{author}{\bibinfo{person}{Sven Mayer}, \bibinfo{person}{Gierad Laput}, {and} \bibinfo{person}{Chris Harrison}.} \bibinfo{year}{2020}\natexlab{}.
\newblock \showarticletitle{Enhancing mobile voice assistants with worldgaze}. In \bibinfo{booktitle}{\emph{Proceedings of the 2020 CHI Conference on Human Factors in Computing Systems}}. \bibinfo{pages}{1--10}.
\newblock


\bibitem[Oudah et~al\mbox{.}(2020)]%
        {oudah2020hand}
\bibfield{author}{\bibinfo{person}{Munir Oudah}, \bibinfo{person}{Ali Al-Naji}, {and} \bibinfo{person}{Javaan Chahl}.} \bibinfo{year}{2020}\natexlab{}.
\newblock \showarticletitle{Hand gesture recognition based on computer vision: a review of techniques}.
\newblock \bibinfo{journal}{\emph{journal of Imaging}} \bibinfo{volume}{6}, \bibinfo{number}{8} (\bibinfo{year}{2020}), \bibinfo{pages}{73}.
\newblock


\bibitem[Rekimoto and Nishimura(2021)]%
        {rekimoto2021derma}
\bibfield{author}{\bibinfo{person}{Jun Rekimoto} {and} \bibinfo{person}{Yu Nishimura}.} \bibinfo{year}{2021}\natexlab{}.
\newblock \showarticletitle{Derma: silent speech interaction using transcutaneous motion sensing}. In \bibinfo{booktitle}{\emph{Proceedings of the Augmented Humans International Conference 2021}}. \bibinfo{pages}{91--100}.
\newblock


\bibitem[Yu et~al\mbox{.}(2024)]%
        {yu2024explainable}
\bibfield{author}{\bibinfo{person}{Mengjie Yu}, \bibinfo{person}{Dustin Harris}, \bibinfo{person}{Ian Jones}, \bibinfo{person}{Ting Zhang}, \bibinfo{person}{Yue Liu}, \bibinfo{person}{Naveen Sendhilnathan}, \bibinfo{person}{Narine Kokhlikyan}, \bibinfo{person}{Fulton Wang}, \bibinfo{person}{Co Tran}, \bibinfo{person}{Jordan~L Livingston}, {et~al\mbox{.}}} \bibinfo{year}{2024}\natexlab{}.
\newblock \showarticletitle{Explainable Interfaces for Rapid Gaze-Based Interactions in Mixed Reality}.
\newblock \bibinfo{journal}{\emph{arXiv preprint arXiv:2404.13777}} (\bibinfo{year}{2024}).
\newblock


\bibitem[Yuan et~al\mbox{.}(2021)]%
        {yuan2021invoking}
\bibfield{author}{\bibinfo{person}{Yuan Yuan}, \bibinfo{person}{Kenneth Mixter}, {and} \bibinfo{person}{Tuan Nguyen}.} \bibinfo{year}{2021}\natexlab{}.
\newblock \bibinfo{title}{Invoking automated assistant function (s) based on detected gesture and gaze}.
\newblock
\newblock
\shownote{US Patent 10,890,969}.


\end{thebibliography}

\appendix

\end{document}